\documentclass[12pt]{iopart}
\usepackage{setstack,iopams}
\usepackage{graphicx}
\newcommand{\bazro}{BaZrO$_3$}
\newcommand{\cazro}{CaZrO$_3$}

\begin{document}

\title[Reactive Force Field for Proton Diffusion in \bazro]
{Reactive Force Field Simulation of Proton Diffusion in \bazro ~using an Empirical Valence Bond Approach}

\author{Paolo Raiteri,$^1$ Julian D Gale$^1$ and Giovanni Bussi$^2$}

\address{$^1$ Nanochemistry Research Institute, Department of Chemistry, Curtin University,
GPO Box 1987, Perth, WA 6845, Australia.}
\eads{\mailto{paolo@ivec.org}, \mailto{julian@ivec.org}}

\address{$^2$ Scuola Internazionale Superiore di Studi Avanzati (SISSA), Via Bonomea 265, 34136 Trieste, Italy}

\begin{abstract}

A new reactive force field to describe proton diffusion within the solid-oxide fuel cell material 
\bazro ~has been derived. Using a quantum mechanical potential energy surface,
the parameters of an interatomic potential model to describe hydroxyl groups within both
pure and yttrium-doped \bazro ~have been determined. Reactivity is then incorporated
through the use of the empirical valence bond model. Molecular dynamics simulations (EVB-MD) have
been performed to explore the diffusion of hydrogen using a stochastic 
thermostat and barostat whose equations are extended to the isostress-isothermal ensemble. 
In the low concentration limit, the presence of yttrium is found not to significantly influence 
the diffusivity of hydrogen, despite the proton having a longer residence time at oxygen adjacent 
to the dopant. This lack of influence is due to the fact that trapping occurs infrequently, 
even when the proton diffuses through octahedra adjacent to the dopant. 
The activation energy for diffusion is found to be 0.42 eV,
in good agreement with experimental values, though the prefactor is slightly underestimated.

\end{abstract}


\submitto{Journal of Physics Condensed Matter}

\clearpage
\section{Introduction}

Solid Oxide Fuel Cells (SOFCs) are an important technology for energy production, including as part of a hydrogen economy. They are of interest for both oxygen and hydrogen ion conduction, where the latter is of particular relevance to trying to achieve a reduction in operating temperatures \cite{Fabbri10,Malavasi10}. Perovskite structured materials have attracted considerable interest for proton conduction, and one of the best candidates is  \bazro. When doped with yttrium on the zirconium B-site, this system achieves one of highest levels of proton conductivity known, combined with other favourable properties, such as good chemical stability \cite{kreuer03}. 

Computational methods offer the opportunity to obtain insights into the atomistic details of proton conduction within materials. There have been numerous theoretical studies to date of alkaline earth metal zirconate-based perovskites, especially those of calcium and barium \cite{munch97,Davies99}. These two systems provide an interesting contrast since while barium zirconate is a cubic perovskite, the calcium analogue exhibits a distorted orthorhombic structure. Both cases have been examined quantum mechanically using density functional theory 
\cite{BilicGale07,Zhong95,Bennett06}. Here barium zirconate is found not to be cubic as it possesses a phonon instability along the R-M edge of the Brillouin zone \cite{BilicGale09}. This represents a failure of present day exchange-correlation functionals within both the Local Density and Generalized Gradient Approximations. Although some density functional studies have reported no distortion from the cubic structure \cite{gomez05}, this has been shown to be a result of the lack of semi-core states in the pseudopotential used to describe barium \cite{BilicGale09}. For both  \cazro ~and  \bazro, the doping and proton conduction pathways have also been studied quantum mechanically. In the former case, this represents a challenging undertaking due to the large number of possible symmetry inequivalent proton hops \cite{BilicGale07}. For  \bazro, the situation is much simpler with proton migration being determined by three motions; two hydroxyl rotations, one about the B-O-B axis and one orthogonal to this, where B represents an octahedral cation, and a jump between two neighbouring oxygens of the same B-site cation along the edge of the
octahedron.

While the study of proton migration in bulk perovskite materials, including those with a single dopant, is readily amenable to quantum mechanical examination, the full mapping of the potential energy landscape becomes challenging as defects are introduced that lower the symmetry, such as grain boundaries, dislocations, and high concentrations of dopants. Hence, there is a need to explore other techniques to examine proton diffusion on larger length scales and for more complex configurations. The issue of length and timescale can be addressed in part by the use of kinetic Monte Carlo using quantum mechanically determined rate constants, as has been performed for  \cazro 
~\cite{BilicGale08}, or via vertex coding, as has been applied to \bazro ~\cite{gomez10}. However, the determination of the rate constants can be time consuming and assumptions have to be made regarding the transferability of values between distinct environments. 

In contrast to quantum mechanical methods, the use of interatomic potentials is readily applicable to large and complex atomistic models of materials. There has already been extensive application of simulation to solid oxide fuel cells using force fields based on a shell model description of the polarisable oxide anion. Here both bulk and defect properties have been investigated, and the thermodynamics of oxygen vacancies, water incorporation and proton trapping by dopants are all accessible. While the activation energies for oxygen diffusion within perovskites can also be calculated from such force fields, the case of proton diffusion is more problematic. Many of the force field simulations of hydrogen incorporation into oxides to date have employed the model for the hydroxyl group developed by Saul \textit{et al} \cite{SaulCatlow} in which this entity is described as a covalent molecule, i.e. the oxygen and hydrogen are connected by a Coulomb-subtracted Morse potential. Because of the form of this potential, the O-H bond cannot be smoothly dissociated and therefore the description is limited to equilibrium configurations involving hydroxyl groups. Although there have been attempts to describe protons using a formal charge model with a Buckingham repulsion \cite{ZeoPots}, as used widely for metal cations, this approach has only met with partial success and does not lead to a broadly applicable model for hydrogen in oxides.

Recently, van Duin \textit{et al} \cite{vanduin08} derived a parameterisation of the reactive force field method, ReaxFF \cite{ReaxFF}, designed for the simulation of proton diffusion in  \bazro. Here the interactions are described as a sum of many different contributions representing terms from covalent bond order, through to van der Waal's and electrostatic energies, based on structure dependent charges. The many parameters were determined by fitting against a database of quantum mechanical potential energy surfaces spanning the component metals, alloys, binary oxides and barium zirconate itself. The resulting model was demonstrated to accurately reproduce the experimental activation energy for proton diffusion during reactive molecular dynamics  \cite{vanduin08}. However, this parameterisation suffers from a notable limitation in that  \bazro ~is predicted to be highly non-cubic as there are six imaginary modes in the phonon spectrum at the $\Gamma$-point \cite{GaleWrightRIMG}. If minimised without symmetry constraints, and from a perturbed initial geometry, a stable configuration is found in which the unit cell is slightly distorted from cubic to triclinic. More seriously, the coordination geometry about zirconium is altered to $C_{2v}$ local symmetry with O-Zr-O bond angles ranging between 74 and 108$^{o}$ instead of all being equal to 90$^{o}$. As a result, many of the distortions observed during proton doping may be the result of symmetry breaking in the unstable bulk configuration, rather than as a result of the defect itself. The consequences for the activation energy of diffusion and pathways are more difficult to assess.

Although it would be possible to refit the parameterisation of the ReaxFF methodology to ensure a cubic ground state for  \bazro, this is a non-trivial undertaking since many of the parameters are strongly correlated. Hence, in the present work our objective is to explore an alternative approach to deriving a reactive force field for proton diffusion in  \bazro ~that is simpler to parameterise. Here we first construct an unreactive force field model that reproduces a combination of quantum mechanical and experimental data for hydrogen in Y-doped barium zirconate. Reactivity is then included through the use of Empirical Valence Bond theory \cite{WarshelWeiss}, to couple together the possible states of the proton within the material, as has been successfully employed to simulate proton transfer in aqueous systems \cite{MSEVB3}. After describing the derivation of the present model, we present results to illustrate its application to proton diffusion in both pure and Y-doped  \bazro.

\section{Methodology}

\subsection{Quantum mechanical calculations}

All quantum mechanical calculations have been performed within the framework of Kohn-Sham density functional theory (DFT) using the SIESTA methodology \cite{SIESTA2} and code for linear-scaling Hamiltonian construction. Here the core electrons and nuclei are represented through the use of non-local norm-conserving pseudopotentials of the modified Troullier-Martins form \cite{TMpseudo}. Small core pseudopotentials have been utilised for Ba, Zr and Y that were generated for valence electron configurations of $5s^{2}5p^{6}$, $4s^{2}4p^{6}4d^{2}$ and $4s^{2}4p^{6}4d^{1}$, respectively (i.e. all metal species are dications). Kohn-Sham states for the valence electrons are expanded as a linear combination of Pseudo Atomic Orbitals. The first basis function for each angular momentum channel is the numerical solution for the isolated atom on a logarithmic grid, subject to radial confinement. Soft-confinement is employed, according to the method of Junquera \textit{et al} \cite{SIESTA1}, with a potential of 100 Ry and an inner radius equal to 0.95 of the outer boundary. The radius of confinement for each orbital is set individually based on the spatial extent of the function. Values of this cut-off for bound states vary between 6.0 Bohr for the semi-core states to 10.0 Bohr for the relatively diffuse 5p levels of Y and Zr. To increase the variational freedom, additional radial functions were generated using the split-norm construct, with a norm of 0.5 for hydrogen and 0.15 for the remaining atoms, as well as including polarisation functions for all atoms (2p, 3d and 4f, for H, O and Ba/Zr/Y, respectively). The cationic metals were described using a basis set of double-zeta polarised quality (DZP), while oxygen and hydrogen were assigned a triple-zeta, doubly polarised (TZ2P) basis. 

In the SIESTA methodology the electron density is expanded in a uniform grid of points in Cartesian space. A kinetic energy cut-off of 600 Ry was used to ensure a high degree of convergence with respect to this auxiliary basis set. The electronic structure was integrated across the Brillouin zone using a Monkhorst-Pack mesh of dimensions controlled by the equivalent real-space cut-off of 12 \AA ~according to the scheme of Moreno and Soler \cite{MorenoSoler}. Geometry optimisations were performed until atomic forces were less than 0.01 eV/\AA, while the cell stresses were considered converged when below 200 bar, though in practice the residual stresses are typically an order of magnitude less than this by the time the internal force tolerance is satisfied. In calculations where a net charge is present in the unit cell, a neutralising background is applied.

Previous DFT calculations for \bazro ~have explored a range of exchange-correlation functionals, including within the localised basis set approach used in the present work \cite{BilicGale09}. All functionals are found to give similar qualitative results in that this material is predicted to be unstable in the cubic form. The main variation is in the magnitude of the lattice parameter obtained. For the current work we employ one of a number of recent GGA functionals that appear to offer superior results for condensed phases. In particular, we choose to use the AM05 functional \cite{AM05}, which has been demonstrated to give good results for lattice parameters, the relative energies of polymorphs \cite{Spagnoli10} and, importantly when considering proton incorporation, even water clusters \cite{AM05water}.

Given the distorted nature of the equilibrium structure of \bazro ~with all exchange-correlation functionals explored to date, this creates an issue with regard to the symmetry of the material. For the experimental cubic system there are only a limited number of symmetry unique pathways for proton migration. If the ground state DFT structure is used then this symmetry is lost and the number of transition states to be located increases significantly. Hence, for reasons of both simplicity and to remain faithful to the experimental reality, we have chosen to work with the cubic cell. Because the phonon instability arises at the zone boundary and not at the $\Gamma$-point, we can suppress the effects of the distortion by using supercells based on odd multiples of the primitive unit cell. Thus we perform calculations for cells of size 1x1x1, 3x3x3 and 5x5x5 in the present work. 

\subsection{Force field derivation}

All force field derivation has been performed with the program GULP \cite{GULP97,GULP3}, using least-squares refinement against quantum mechanical structures and energy differences. Experimental mechanical properties were also included in the fits to bulk BaO and \bazro ~to better constrain the curvature of the potential energy surface. The relaxed fitting algorithm \cite{GULPFit} has been used in which the structure is optimised at every point of the fit so that the change in structure can be used as the observable, rather than the forces and stresses.

Although the shell model has been widely used to simulate oxide phases in force field studies, here we chose a rigid ion description for Y-doped \bazro. The main reason for this choice is the desire to be able to perform efficient molecular dynamics simulations of proton diffusion and so the incorporation of a shell model would increase the cost by approximately an order of magnitude, regardless of whether using an adiabatic or fictitious mass Lagrangian algorithm. Furthermore, because \bazro ~is a cubic perovskite, the dipolar oxygen polarisability must be kept low otherwise the material would undergo a transformation with a rhombohedral distortion. 

Formal charges are adopted for Ba, Zr, Y and O as an oxide ion, while the short-range interaction between these species is given by the Buckingham potential:

\begin{equation}
U_{ij}(r)=A_{ij} \exp \bigg(-\frac{r}{\rho_{ij}}\bigg) - \frac{C_{ij}}{r^6}
\end{equation}

For all interactions we have decided to fix the value of the dispersion coefficient, $C_{ij}$, to be zero. In the case of the cations this is readily justified, with the possible exception of Ba, due to the low polarisability of the species, while for the oxygen-oxygen interaction the relatively uniform cohesive attraction in the high symmetry cubic structure could be subsumed into other parameters with no loss of quality. Since the use of formal charges probably overestimates the binding contribution from the electrostatic interactions, no further attractive contribution is required from the dispersion forces. Furthermore, given that a fit is being performed to the results of a GGA calculation, there is no long-range van der Waal's contribution to the underlying energy surface.

Fitting of the metal-oxygen interaction parameters was performed sequentially to several structures. Firstly, the Ba-O interaction was fitted to barium oxide alone and then held fixed during the fitting of Zr-O to barium zirconate. Secondly, having determined the metal-oxide potentials, the metal-hydroxide potentials were fitted to data for an excess proton in pure \bazro ~in order to constrain this interaction in the absence of a dopant. Finally, the remaining metal-oxygen potentials involving yttrium were determined based on the quantum mechanical results for a 3x3x3 supercell containing a single Y/H defect pair in a range of different configurations, as well as for the extreme
composition of HBaYO$_{3}$. Here the bond lengths, hydrogen bonding distances and relative energies of the configurations for the supercell were selected as observables.

To describe the hydroxyl group, we adopt a similar approach to previous works by using a molecular mechanics description. Here the oxygen and hydrogen possess partial charges that are constrained to sum to the formal overall charge of -1, while being Coulombically screened from each other. A harmonic potential is used for the short-range intramolecular interaction, while a purely repulsive Lennard-Jones potential is used to capture the intermolecular repulsion between hydrogen and oxygen. The charge distribution within the hydroxyl group and intermolecular terms were fitted against the quantum mechanical structures and activation energies for hydroxyl group rotation within bulk \bazro ~in the absence of the dopant to ensure these two important contributions to proton migration are accurately represented, at least within the limits of the underlying DFT functional. In order to better reproduce the relative energetics of different Y/H defect configurations it was necessary to distinguish between oxide ions in the Zr-O-Zr and Y-O-Zr environments, which will be labelled as O2 and O3 respectively, when interacting with a neighbouring hydroxyl group.  A larger Lennard-Jones repulsion was found to be necessary between an adjacent hydroxyl group and the oxygen coordinated to yttrium (O3).

In order to allow for proton diffusion, reactivity is included using the Empirical Valence Bond (EVB) approach. Here two states, 1 and 2, are coupled by a Hamiltonian:

\begin{equation*} 
\mathbf{H} = 
\left[\begin{array}{ccc}
H_{11} & H_{12}  \\
H_{21} & H_{22}  \\ 
\end{array} \right] 
\end{equation*}

The on-diagonal matrix elements, $H_{11}$ and $H_{22}$, are simply given by the conventional force field energies of the states being coupled, while the off-diagonal elements control the mixing between the states. In the present scenario, the two states represent the proton being covalently bonded to two different oxygens at the same time in which the force field attributes of oxide and hydroxyl oxygen are interchanged. This approach has been widely used for proton transfer, and other, reactions during simulations, especially in liquid water, where models for both an excess proton and the hydroxyl anion have been 
developed \cite{MSEVB3,MSEVBborgis,MSEVBmod,MSEVBoh,MSEVBoh2}. 

In the case of an excess proton in water, it is well documented that the EVB scheme must be extended to the multi-state (MSEVB) variant. Due to the nature of the Grotthuss mechanism, the proton cannot be considered strongly localised on two water molecules alone and so both the first and second solvation shells often need to be included explicitly in the Hamiltonian. In the specific case of perovskite fuel cells it is highly improbable that a chain of hydroxyl groups will exist such that multi-state coupling is required. Firstly, the concentration of dopants, and therefore protons, is usually sufficiently low that two hydrogens will rarely be located on nearest neighbour oxygens. Secondly, even in a situation where this did occur, the geometric constraints are much greater in the solid-state than in the liquid, and so the ability to attain a favourable configuration for formation of a coupled chain of hydroxyls is far more limited. Hence, even if the case of a single proton coupling to more than two oxygens is contemplated, this is unlikely to occur. 
Furthermore, proton migration between oxygens of the same octahedron is found to occur along the edges and avoid the faces where three oxygens would become simultaneously coupled to the same hydrogen.

Within the EVB method it is necessary to define the off-diagonal matrix elements that control the strength of coupling between a hydroxyl group and the oxygen to which it is most strongly hydrogen bonded. Several functional forms have been proposed for the coupling elements in EVB in terms of the reaction coordinate for proton transfer. The important properties that they must possess are a smooth decay to zero as the distance between the groups tends to infinity and a derivative of zero with respect to the reaction coordinate when the O-H..O group is symmetric in order to avoid an artificial cusp. Similarly, there are numerous choices for the reaction coordinate. Here we take the simple choice of a reaction coordinate, $Q$, that is defined as the difference between the two O-H bond lengths. We then define the coupling matrix element using a Gaussian form:

\begin{equation}
H_{ij}(r)=\lambda_{ij} \exp(-\zeta_{ij}.Q^{2})
\end{equation}

The two parameters in this expression have been determined as follows. Firstly, the exponent for decay, $\zeta$, was selected
such that the coupling constant was negligible at the equilibrium geometry for a hydroxyl defect in \bazro. This allows the fitting of the equilibrium structures and the reactive configurations to be separated. Secondly, the prefactor for the coupling, $\lambda$, is just the difference between the force field activation energy at the transition state for proton migration and the value calculated quantum mechanically. Hence, this quantity was determined for the symmetric transition state of a proton 
migrating in pure \bazro. While potentially the coupling matrix elements could depend on the local environment of 
the hydroxyl group, in the spirit of force field transferability we make the assumption that the coupling matrix elements are independent of
the nature of the surrounding metal cations, though of course the reactivity will differ still due to the underlying 
force field contribution. 

As a final refinement to the present model, we also include a self-energy contribution for the
situation when the hydroxyl group is coordinated to Y in order to obtain the correct relative energy
in comparison to this anion being remote to the dopant. While in principle this difference could be
captured by the interaction between yttrium and the hydroxyl group being assigned a different
short-ranged potential, this factor is simpler to correct for using the self-energy.

\subsection{Molecular dynamics simulations}

Having determined the reactive force field, we are in a position to simulate proton diffusion within \bazro.
Given the low activation energy for migration and elevated temperature of operation of fuel cells, it is 
appropriate to simulate diffusion using molecular dynamics. Because the majority of molecular dynamics
codes do not allow for the Empirical Valence Bond model (EVB-MD), we have written a new program to handle this
known as ReaxMD. In this code we have implemented a flexible pressure/stress control, which
allows one to sample configurations from the NST ensemble. This is
probably not necessary for the present application, since the
simulation box is large and the system is stiff, but could be crucial
in the simulation of softer materials or of smaller boxes. Below we describe in detail the employed algorithm.

The equations of motion that we employ are a modification suitable for flexible cells
of those derived in \cite{bussi09jcp}. In that work, several approaches were compared and it was found that
an optimal choice is to combine a standard $NPH$ scheme with a stochastic velocity rescaling thermostat
\cite{bussi07jcp} so as to sample the $NPT$ ensemble. The stochastic velocity rescaling combines
the ergodicity of stochastic schemes with the efficiency of global thermostats \cite{bussi08cpc}, while also retaining the
possibility of defining a conserved energy to control the time step discretisation errors.

Several schemes are available for flexible cell sampling, $NSH$, see for example \cite{parrinello80prl, parrinello81jap, martyna94jcp}. Here we build our scheme in a fashion similar to that
of \cite{martyna94jcp}, but introduce corrections so as to control exactly
the sampled ensemble. Defining
$\mathbf{r}_i$ and
$\mathbf{p}_i$ the positions and momenta of the $i$-th particle, respectively,
$\bi{\mathbf{h}}$ as the matrix of the lattice vectors, written as columns
$\bi{a}=(h_{11},h_{21},h_{31})$,
$\bi{b}=(h_{12},h_{22},h_{32})$,
$\bi{c}=(h_{13},h_{23},h_{33})$, 
and $\bi{\mathbf{p}}$ as the matrix of the associated momenta,
our equations of motion read:
\begin{equation}
\begin{array}{rl}
\label{eq:nph}
\displaystyle\dot{\mathbf{r}}_i&= \displaystyle\frac{\mathbf{p}_i}{m_i} + \frac{\bi{\mathbf{p}}}{W} \mathbf{r}_i, \\
\label{eq:nph-p}
\displaystyle\dot{\mathbf{p}}_i&= \displaystyle\mathbf{f}_i - \frac{\bi{\mathbf{p}}}{W} \mathbf{p}_i, \\
\label{eq:nph-pcell}
\displaystyle\dot{\bi{\mathbf{p}}}&= \displaystyle\Big[\Big(\bi{\mathbf{P}}_{\mathrm{int}}-
\bi{\mathbf{\Pi}}_{\mathrm{ext}}\Big)V
\displaystyle +2k_BT\bi{\mathbf{L}}\Big] \\
\dot{\bi{\mathbf{h}}}&= \displaystyle\frac{\bi{\mathbf{p}}
\bi{\mathbf{h}}}{W}
\end{array}
\end{equation}
Here $W$ is the barostat mass, $m_i$ the mass of $i$-th particle, 
$V=\det{[\bi{\mathbf{h}}]}$,
the volume of the simulation cell,
$\bi{\mathbf{L}}$ is a diagonal tensor with 
$(L)_{\alpha\alpha}=4-\alpha$, $k_B$ is the Boltzmann constant, $T$ the temperature,
$\bi{\mathbf{P}}_{\mathrm{int}}$ is the internal stress and
$\bi{\mathbf{\Pi}}_{\mathrm{ext}}$ is;

\begin{equation}
\bi{\mathbf{\Pi}}_{\mathrm{ext}} = 
\frac{V_0}{V} \Bigg[
\bi{\mathbf{h}}\bi{\mathbf{h}}_0^{-1}
\bi{\mathbf{P}}_{\mathrm{ext}}
(\bi{\mathbf{h}}_0^T)^{-1}\bi{\mathbf{h}}^{T}
\Bigg]
\end{equation}
where $\bi{\mathbf{P}}_{\mathrm{ext}}$ is the external load and has to be symmetric
to avoid any unphysical rotations of the cell.
The introduction of a reference cell, $\bi{\mathbf{h}}_0$,
is needed here to take into account the internal elastic energy accumulated in the simulation 
cell~\cite{parrinello80prl, parrinello81jap}.
In order to avoid any cell rotations we fix the reference system so that 
$\bi{\mathbf{h}}$ is upper half triangular
and build an initial $\bi{\mathbf{p}}$ that is also upper half triangular.
This is equivalent to removing the rotational degrees of freedom of the system, as discussed 
in \cite{martyna94jcp}.

By inspection, one can prove that the equations of motion (\ref{eq:nph}), besides the total momentum
and the position of the centre of mass, also conserve the following quantity;
\begin{equation}
\label{eq:conservednph}
H=K+U+U_{el}
+\frac{\mathrm{Tr}[\bi{\mathbf{p}}^T\bi{\mathbf{p}}]}{2W}
-2k_BT \sum_{i=\alpha}^3 (4-\alpha)\log (h)_{\alpha\alpha}
\end{equation}
where $K$ is the total kinetic energy of the particles, $U$ is their potential energy,
and $U_{el}$ is the elastic energy stored in the system~\cite{parrinello81jap};
\begin{equation}
U_{el}=V_0\mathrm{Tr}\Big[\bi{\mathbf{P}}_{\mathrm{ext}} 
\bi{\epsilon}\Big]
\end{equation}
where $V_0$ is the volume of the reference cell and
$\bi{\epsilon}$ is the strain tensor, which can be calculated as:
\begin{equation}
\bi{\epsilon} = \frac{1}{2} 
\bigg((\bi{\mathbf{h_0}}^T)^{-1}\bi{\mathbf{h}}^T
\bi{\mathbf{h}}\bi{\mathbf{h_0}}^{-1}
-\bi{\mathbf{I}}\bigg).
\end{equation}
By a careful calculation one can show that the Jacobian
of the equations of motion~(\ref{eq:nph}) is proportional to
$(h)_{11}^3(h)_{22}^2(h)_{33}^1$.
This result is different from that of Martyna {\it et al}~\cite{martyna94jcp} since
we are not including a thermostat at this stage and
we are fully taking into account the constraints on the cell shape (upper half triangular)
when computing the Jacobian.

Combining the above information, one can state that ~(\ref{eq:nph}) produces
the ensemble;
\begin{equation}
\label{eq-nph-probability}
\begin{array}{l}
\displaystyle\mathcal{P}_{NSH}(p,r,\bi{\mathbf{h}},\bi{\mathbf{p}})
\mathrm{d}p
\mathrm{d}r 
\mathrm{d}^6 \bi{\mathbf{h}} 
\mathrm{d}^6 \bi{\mathbf{p}}
\propto \\
\displaystyle (h)_{11}^{-3}(h)_{22}^{-2}(h)_{33}^{-1}
\delta\left(\sum_im_i\mathbf{r}_i\right)
\times
\delta\left(\sum_i\mathbf{p}_i\right)
\delta\left( H-H_0 \right)
\mathrm{d}p
\mathrm{d}r 
\mathrm{d}^6 \bi{\mathbf{h}} 
\mathrm{d}^6 \bi{\mathbf{p}}
\end{array}
\end{equation}
where $H_0$ is the initial value of $H$. Here we write $\mathbf{d}^6 \bi{\mathbf{h}}$
as a reminder that we are only sampling the 6-dimensional space of all the
possible upper half triangular cell matrices. The change of measure
can be performed by considering the relation
$(h)_{11}^2(h)_{22} \mathbf{d}^6 \bi{\mathbf{h}}
\propto
\mathbf{d}^9 \bi{\mathbf{h}}$.

When combined with a thermostat, the resulting ensemble is:
\begin{equation}
\label{eq-npt-probability}
\begin{array}{l}
\displaystyle
\mathcal{P}_{NST}(p,r,\bi{\mathbf{h}},\bi{\mathbf{p}})
\mathrm{d}p
\mathrm{d}r 
\mathrm{d}^9 \bi{\mathbf{h}} 
\mathrm{d}^6 \bi{\mathbf{p}}
\propto
V\delta\left(\sum_im_i\mathbf{r}_i\right) \\
\displaystyle
\times
\delta\left(\sum_i\mathbf{p}_i\right)
\exp \Bigg[
-\frac{K+U
+U_{el}
+\frac{\mathrm{Tr}[\bi{\mathbf{p}}^T\bi{\mathbf{p}}]}{2W}}{k_BT}
\Bigg]
\mathrm{d}p
\mathrm{d}r 
\mathrm{d}^9 \bi{\mathbf{h}} 
\mathrm{d}^6 \bi{\mathbf{p}}
\end{array}
\end{equation}
This ensemble is the usual $NST$ ensemble, with an extra
$V$ term that compensates for the constraint on the 
centre of mass of the system.
In this way, the expectation value of any observable depending only on the
inter-particle distance or cell shape is unaffected by the centre of mass constraint
and is identical to what would be obtained from a Monte Carlo simulation.
For a more detailed discussion of this term, see \cite{bussi09jcp}.
The apparently asymmetric contribution in
~(\ref{eq:nph-pcell}) is indeed necessary to explore the correct ensemble. 
We have tested our implementation on a small model system and verified that, for an external
isotropic pressure, this correction is needed to recover invariance with respect
to permutation of the lattice vectors.

In order to numerically forward integrate the equations of motion (\ref{eq:nph}) for a 
time interval, $\Delta t$,
we use a scheme based on the Trotter-decomposition~\cite{trotter1959,
tuckerman92jcp}, which is a straightforward extension of the algorithm used
for the isotropic, $NPT$, case~\cite{bussi09jcp}:
\begin{itemize}
\item Evolve thermostat equations ($T$-step) for $\Delta t/2$.
\item Evolve velocities ($P$-step for $\Delta t/2$.
\item Evolve positions and velocities  ($R$-step) for $\Delta t$.
\item Evolve velocities ($P$-step) for $\Delta t/2$.
\item Evolve thermostat equations ($T$-step) for $\Delta t/2$.
\end{itemize}
The $T$-step rescales the particle velocities and the
cell momenta with the same factor, which is obtained
through a stochastic procedure \cite{bussi09jcp, bussi07jcp}. Note that
here the total kinetic energy now includes also the cell kinetic energy, 
$\mathrm{Tr}[\bi{\mathbf{p}}^T\bi{\mathbf{p}}] / 2W$,
and that the number of degrees of freedom is augmented so as to also account for
the 6 cell degrees of freedom.

The $P$-step is the evolution of the atoms' and cell momenta and can be written as;

\begin{equation}
\begin{array}{rl}
\displaystyle\mathbf{p}_i\left(t+\frac{\Delta t}{2}\right) &= 
\displaystyle \mathbf{p}_i(t)+\mathbf{f}_i(t)\frac{\Delta t}{2}\\
\displaystyle(p)_{\alpha\beta}\left(t+\frac{\Delta t}{2}\right)
&=\displaystyle
(p)_{\alpha\beta}(t)+
\Big[V(\bi{\mathbf{P}}_{\mathrm{int}}-
\bi{\mathbf{\Pi}} _{\mathrm{ext}})
+2k_BT\bi{\mathbf{L}}\Big]
\frac{\Delta t}{2} \\
&\displaystyle
+\sum_i\frac{
(\mathbf{f}_i(t))_{\alpha} (\mathbf{p}_i(t))_{\beta}
+
(\mathbf{p}_i(t))_{\alpha} (\mathbf{f}_i(t))_{\beta}
}{2m_i}
\left( \frac{\Delta t}{2} \right)^{2} \\
& \displaystyle +\frac{
(\mathbf{f}_i(t))_{\alpha} (\mathbf{f}_i(t))_{\beta}
}{3m_i}
\left( \frac{\Delta t}{2} \right)^{3}
\end{array}
\end{equation}
where the last two terms on the right hand side represent the time propagation of the 
kinetic contribution to the stress tensor.
The instantaneous internal strain, $\bi{\mathbf{P}}_{\mathrm{int}}$, is calculated here as;
\begin{equation}
\begin{array}{rl}
(P_{\mathrm{int}})_{\alpha\beta}(t)&= \displaystyle
\frac{1}{V}\left[
\sum_i\frac{(\mathbf{p}_i)_{\alpha}(\mathbf{p}_i)_{\beta}}{m_i}+
\sum_{i>j}(\mathbf{f}_{ij})_{\alpha} (\mathbf{r}_{ij})_{\beta}
\right]
-(\bi{U}'\bi{\mathbf{h}}^T)_{\alpha\beta} \\
(U')_{\alpha\beta}&= \displaystyle\frac{\partial U(\mathbf{r},\bi{\mathbf{h}})}{\partial(h)_{\alpha\beta}}
\end{array}
\end{equation} 
and setting to zero the terms below the diagonal:
\begin{equation}
(P_{\mathrm{int}})_{\alpha\beta}=0~\mathrm{for}~\alpha > \beta
\end{equation}

The $Q$-step is the evolution of the atoms' positions and cell vectors and can be written as:
\begin{equation}
\begin{array}{rl}
\mathbf{r}_i(t+\Delta t)&= \displaystyle
e^{\frac{\bi{\mathbf{p}}(t)}{W}\Delta t}
\mathbf{r}_i(t)
+
\left[\frac{\bi{\mathbf{p}}(t)}{W}\right]^{-1}
\left[
\frac{e^{\frac{\bi{\mathbf{p}}(t)}{W}\Delta t}
-e^{-\frac{\bi{\mathbf{p}}(t)}{W}\Delta t}}{2}
\right]\frac{\mathbf{p}_i(t)}{m_i}
\\
\mathbf{p}_i(t+\Delta t)&= \displaystyle
e^{-\frac{\bi{\mathbf{p}}(t)}{W}\Delta t}
\mathbf{p}_i(t)
\\
\bi{\mathbf{h}}(t+\Delta t)&= \displaystyle
e^{\frac{\bi{\mathbf{p}}(t)}{W}\Delta t}
\bi{\mathbf{h}}(t).
\end{array}
\end{equation}
The matrix $e^{\frac{\bi{\mathbf{p}}(t)}{W}\Delta t}$ is obtained
from $\frac{\bi{\mathbf{p}}(t)}{W}\Delta t$ by means of the
Crank-Nicolson expansion truncated at the 5$^{th}$ order~\cite{crank47}.

When a finite time step is adopted, sampling errors are inevitably introduced.
To keep them under control,
one can monitor the drift of an effective energy that quantifies the detailed balance violation
\cite{bussi07jcp, Bussi2007pre}.
Also in the present case, the effective energy drift is computed
by book-keeping the total kinetic energy increment given by the
thermostat and subtracting it from (\ref{eq:conservednph}).

All molecular dynamics simulations have been performed with a stochastic thermostat and the above new barostat
with relaxation times of 0.1 ps and 1 ps, respectively. All simulations were 1 ns long, unless otherwise stated, 
with a 1 fs time step that led to perfect energy conservation for non-reactive trajectories, while for the reactive ones
at high temperature a slight drift was observed (~3 eV out of a total absolute magnitude of ~31200 eV,
which corresponds to a 0.01\% variation, or a drift of ~1 meV per degree of freedom per ns). For every temperature,
12 runs independent runs were performed, employing different starting configurations and random velocities.

\section{Results and discussions}

\subsection{Quantum mechanical calculations}

Optimisations of both BaO and \bazro ~have been performed using the AM05 exchange-correlation functional.
As shown in Table \ref{bulk_results}, the present results show very good agreement with the experimental cell parameters,
especially for \bazro. While the many previous studies of this material have typically exhibited the systematic 
overestimation/underestimation of the lattice parameter associated with use of a GGA/LDA functional, 
the AM05 functional appears to be superior for the present structures. We note that the inclusion of the 
f polarisation functions
for Ba and Zr are important here, in that the lattice parameter increases to 4.210 \AA ~from 4.1898 \AA ~in the 
absence of these. For BaO, the error in the lattice parameter for AM05 is larger, but still only 0.3 \%. 

There have been several previous quantum mechanical studies of barium zirconate and the results found
in the present work are consistent with the more recent of these \cite{BilicGale09} in finding that the structure is phonon stable
at the $\Gamma$-point, but exhibits instabilities elsewhere in the Brillouin zone when cubic. As a result, we have
performed defect calculations at constant volume and using odd multiples of the primitive cell in order to 
suppress the influence of these unstable modes. 

Before considering the doping of \bazro, we first contemplate the incorporation of a proton in a single unit cell
of pure material. Here a charge neutralising background is applied to compensate for the positive charge
of the proton. There is only one symmetry unique structure for the proton in this cell in which the hydroxyl 
group lies at a small angle away from being parallel to a lattice vector and directed partly towards a neighbouring
oxide ion in the same plane as the common zirconium ion. From this site, there are two distinct types of 
activated motion that can occur in which the hydroxyl rotates about two orthogonal axes, one leading to 
a new intermolecular hydrogen bond to an oxygen within the same octahedron and the other leading to 
interaction with a different octahedron. Of these two processes, the first is the most energetic with
a computed activation energy of 0.246 eV. 

In addition to computing the pathways for hydroxyl group reorientation, it is also possible to calculate the
barrier to proton transfer between two adjacent oxygens. By exploiting the symmetry of the transition 
state, this configuration can be readily located via a constrained minimisation. Here a low barrier of 
0.0375 eV is computed, suggesting that proton exchange between neighbouring oxygens may be 
faster than rotational reorientation, though we will discuss a number of caveats on this below.

The activation energies for proton migration in regions away from the dopant have also been calculated
by Bj\"{o}rketun \textit{et al.} \cite{bjoerketun07} using the PW91 functional \cite{pw91}. Here they found 
barriers of 0.16-0.18 and 0.20 eV
for rotation and transfer, respectively. Gomez \textit{et al.} \cite{gomez05} have also computed the same quantities, with the same functional,
but obtain slightly different values of 0.14 and 0.25 eV. 
Qualitatively both of these sets appear to be different to the present work in
that the order is reversed, i.e. rotation would be easier than proton transfer. One of the key differences between
the literature studies and the present work is the functional used and thereby the lattice parameter 
for \bazro. Using the same
computational conditions, we have computed the variation of the rotational energy barrier by increasing 
and decreasing the lattice parameter. For values of 4.0 and 4.4 \AA ~we obtain values of 0.513 and 0.170 eV,
respectively, thus illustrating the strong dependence on lattice parameter. Consequently, the lower value
for this quantity found in earlier work is entirely consistent with their larger lattice constant of 4.25 \AA.
Conversely, the activation energy for the proton transfer reaction is found to 
increase rapidly with increasing lattice parameter, again in line with the difference between the studies.

A study of proton migration in Y-doped \bazro ~by Merinov and Goddard III \cite{Merinov09} also examines the same processes,
this time using calculations based on the PBE functional \cite{pbe}. Given that both PBE and PW91 
are GGA functionals,
one might be expecting them to yield similar results for the present material (though we note that this
is not always the case \cite{pbepw91differ}).
However, a higher activation barrier for proton transfer of 0.48 eV is quoted. In this work they also note the
strong sensitivity of the barrier to the O-O distance along the edge of the octahedron, but there is no
explicit information given regarding the optimised lattice parameter. Thus it is difficult to explain the
apparent discrepancy between this work and the earlier works in the field based on this. However, a recent study by
Gomez \textit{et al.} \cite{gomez10} using the PBE functional also obtains proton transfer barriers of
0.5-0.6 eV away from the dopant. An even higher activation energy for proton transfer of 0.69 eV has been determined for \bazro ~by M\"{u}nch \textit{et al.} \cite{munch97}. 
However, this value was computed within the Local Density Approximation and so systematic
overbinding is to be expected. 

Given the self-interaction errors inherent in GGA Kohn-Sham theory, any values for proton migration
activation energies are likely to be an
underestimate, but the higher value found in the present work relative to Bj\"{o}rketun \textit{et al.} is arguably 
preferable on the basis of the more accurate lattice parameter. The fact that Merinov and Goddard III
obtain closer agreement with experiment may not be indicative of a better result, given the expectation
that Kohn-Sham theory with a GGA functional should not be perfect. Indeed, without considering the
influence of thermal fluctuations on the system one might not expect to agree with high temperature
experimental data. 

An additional reason for the much lower barrier to migration in the present work, 
as compared to those of Merinov and Goddard III or Gomez \textit{et al.} relates to the supercell dimensions.
In both of the aforementioned studies a 2x2x2 supercell was employed, which allows the barium zirconate
structure to distort due to the imaginary phonons at the zone boundary, whereas in our work we 
restrict this by using odd numbered supercells. Therefore, in the higher literature values there is a
contribution to the activation energy from reorientation of the octahedra, which is absent in the 
current study, and we presume in the experimental system that is actually cubic.  Furthermore, the low
barrier here is for a single cell, rather than a supercell and so image interactions may contribute to lowering
the magnitude. While this implies that the present value is not an accurate estimate of the barrier for 
proton migration in the low concentration limit, the objective here is only to provide a potential energy
surface to train a force field against, and so provided the fit is performed for the same configuration
this image interaction is not an issue. The barrier height for more dilute configurations can then be
determined subsequently using the force field instead. 

\begin{figure}[htbp]
   \centering
   \includegraphics[width=10cm]{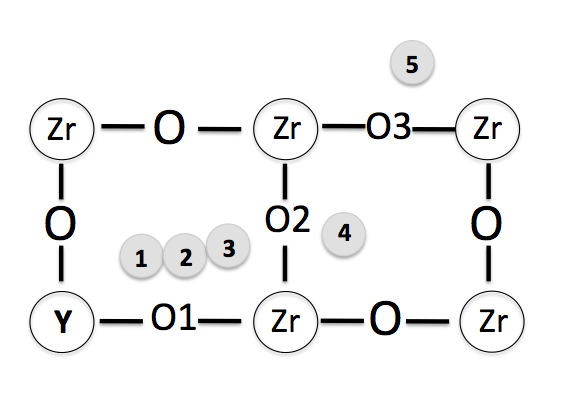} 
   \caption{Schematic illustration of the five proton configurations examined relative to the
   position of yttrium. The numbered sites 1 to 5 refer to the configurations in Table \ref{yoh_energies}.
   Here the hydrogens in positions 1 and 2 are bonded to O1, positions 3 and 4 to O2, while position 5
   is bound to an oxygen one conventional lattice parameter behind O3 into the plane of the figure.}
   \label{fig:scheme}
\end{figure}

Next we turn to consider the question of proton-dopant binding for Y in \bazro. Here we have examined
five different proton binding configurations, shown schematically in Figure \ref{fig:scheme}. In the present work, 
there are two distinct 
minima for a proton bound to the same oxygen within the plane of each permutation of pairs of
lattice vectors (i.e. the OH group is not orthogonal to each B1-O-B2 direction, where B1/B2 = Y or Zr, 
but is tilted slightly towards B1 or B2). For this reason, there are a greater number of minima here than 
in the work of Bj\"{o}rketun \textit{et al.}. The most stable state is for the proton to be bound to an oxygen coordinated
to yttrium, with the hydroxyl tilted to interact with another oxygen of the Y octahedron. In Table \ref{yoh_energies}
the energies
of all configurations relative to this state are presented. Earlier DFT calculations suggest that the proton 
prefers to bind to an oxygen between two Zr ions, rather than adjacent to Y, albeit with a negligible difference
of 0.01 eV. Here we find that same magnitude of energy difference, but with the opposite sign. In a calculation
where the polarisation functions are omitted on Ba and Zr, leading to a large cell parameter, this ordering is
again reversed, suggesting that this value is also slightly sensitive to cell volume. Given that the energy
differences are of order of thermal energies per degree of freedom at ambient conditions, it appears that
the two sites are effectively isoenergetic. 

Comparing the energy difference between the most stable proton binding site and that most remote from
yttrium within the supercell provides an estimate for the trapping energy of -0.237 eV. Stokes and Islam \cite{stokes10}
have computed this quantity based on shell model potentials and obtained a value that appears to be similar,
but marginally more exothermic (value is estimated from Figure 3 in their work). As they note, there are
no corresponding experimental values for exactly this system, though the value lies within the range of -0.2 to
-0.4 eV measured for other materials. Although a recent neutron spin-echo study \cite{Karlsson2010} has examined the proton dynamics
in 10\% ~Y-doped \bazro, no value for the trapping energy is quoted.

\subsection{Force field}

The quantum mechanical information from the previous section has been used as the basis of a force field
derivation for Y-doped \bazro, as described in the methods section. The final parameters are given in Table
\ref{forcefield}. Formal-charged shell model parameters already exist for \bazro ~\cite{stokes10}. 
Although derived independently
of the work of Stokes and Islam, it is reassuring that the resulting Ba-O potential is very similar, as is the 
predicted lattice parameter for this material, though they were fitted to different values. In the present work
the repulsive Zr-O potential is a little steeper, but this may reflect the absence of the oxygen-oxygen attractive
contribution. When fitting the interaction between the metal cations and the hydroxyl oxygen it was generally
found to be sufficient to only vary the $A$ term, rather than the $\rho$ value as well. 

The final results for the fitted force field are compared against the quantum mechanical data, experiment,
and the results of the ReaxFF force field for the bulk properties in Table \ref{bulk_results}. Unsurprisingly, the 
AM05 structures of both BaO and \bazro ~are perfectly reproduced as they were heavily weighted in the fit. 
In contrast, the ReaxFF parameterisation overestimates the lattice parameters for both systems
when constrained to be cubic. With the smaller unit cell dimensions and formal charges, the present force field
overestimates the hardness of  \bazro, though it accurately captures the on-diagonal elastic constant
component, $C_{11}$. Because of the high symmetry cubic structure, even inclusion of a dipolar shell model would
not soften the structure. The symmetry of the Born effective charge tensor for oxygen indicates that a 
quadrupolar or higher polarisability model would be required to improve the description of the curvature.
Conversely, the ReaxFF model predicts mechanical properties for \bazro ~that are too soft. This is almost
certainly due to the lower charges of +1.66 and +1.79 for Ba and Zr, respectively, generated by the 
electronegativity equalisation procedure.  
Although the hardness is a potential weakness of the present model, it should have minimal impact on the proton 
conductivity unless the pressure dependence is examined. 

Having determined the force field parameters required to simulate \bazro, it is now possible to validate 
our algorithm for the integration of the equations of motion (\ref{eq:nph}) using this specific case. To do this,
we compare the elastic constants of \bazro ~as computed from the cell 
fluctuations \cite{parrinello81jap} recorded during a molecular dynamics run at constant 
temperature and pressure against those calculated by lattice dynamics using analytical second derivatives,
as implemented within the program GULP \cite{GULP3}. The elastic constants are determined from the ensemble
average of the strain correlations using the methodology of Parrinello and Rahman \cite{ParrinelloElastic}.
A set of runs were performed with different thermostat and and barostat relaxation times, hereafter indicated
as $\tau_t$ and $\tau_b$, respectively. As a further check we also 
calculated the same properties using a Langevin thermostat with different values of the friction coefficient.
We performed 1~ns long $NST$ runs at 300 K and note that for $\tau_b\ge10~ps$ the cell oscillations were 
extremely slow; thus the simulation time was insufficient to achieve complete convergence of the elastic
constants. On the other hand, $\tau_b<$ 0.1~ps leads to instabilities in the integration of the equation, 
preventing good energy conservation from occurring with a 1~fs timestep, while for intermediate values the calculated 
elastic constants appear to be quite insensitive to the thermostat and barostat relaxation times.

The calculated elastic constants from the molecular dynamics simulations are compared against those
from lattice dynamics in Table \ref{tab:booktabs}. Here the results obtained using the stochastic and Langevin
thermostats are essentially equivalent to within the statistical uncertainty, with the possible exception of 
$C_{12}$. In contrast, there is a definite discrepancy between the values of $C_{11}$ obtained using
molecular dynamics and that obtained from lattice dynamics for the energy minimised unit cell. However, 
this difference is a result of finite temperature effects; as the unit cell of \bazro ~expands, so the value of 
$C_{11}$ decreases. If the lattice dynamics calculation is performed using the average cell dimensions
found during the molecular dynamics trajectory at 300 K, then the agreement between the techniques
is significantly improved. We note that the thermal expansion is considerably overestimated by molecular
dynamics below the Debye temperature due to the lack of vibrational quantisation. Hence, if a free
minimisation is used to determine the thermal expansion, then the reduction in $C_{11}$ would be 
considerably less.

Turning now to consider the energetics of proton defects in Y-doped \bazro, the results from the present
force field are compared against the reference quantum mechanical data in Table \ref{yoh_energies}. While the
long-range limit of the proton trapping energy is explicitly fitted via the self-energy of the hydroxyl group
when coordinated to yttrium, the relative energies of intermediate configurations are also well reproduced.
The greatest difficulty is in getting the energetic balance right between subtly different configurations where
the proton is close to yttrium. Although the percentage errors are high for configurations 2 and 3, the absolute
error in both cases is less than 0.05 eV, which is of the order of ambient thermal energy per degree of freedom 
under the conditions at which fuel cells typically operate. Importantly, the ordering of the energies is correct
and the trapping energy matches the quantum mechanical result. Given that the largest thermodynamic
difference to be overcome for proton diffusion away from the dopant is correct, it is likely that the small errors
in the proton dynamics close to yttrium will not prove critical except at dopant concentrations higher than those
used experimentally. 

For the final step in the force field parameterisation it is necessary to consider the proton transfer event
between two adjacent oxygens. Here we have fitted the parameters of the coupling matrix elements of
the EVB for the case of pure \bazro. The resulting values of the prefactor ($\lambda$) and exponent ($\zeta$) are 0.7998 eV and 16.0 \AA $^{-2}$,
respectively. With these values the density functional barrier to proton transfer is reproduced. 
The barrier for hydroxyl rotation as given by the force field is not guaranteed
to be exactly equivalent to the density functional value since the hydroxyl model has to balance 
many constraints. However, the computed value of 0.231 eV is in good agreement with the first principles value of 0.246 eV.

\subsection{Proton diffusion}

\begin{figure}[htbp]
   \centering
   \includegraphics[width=10cm]{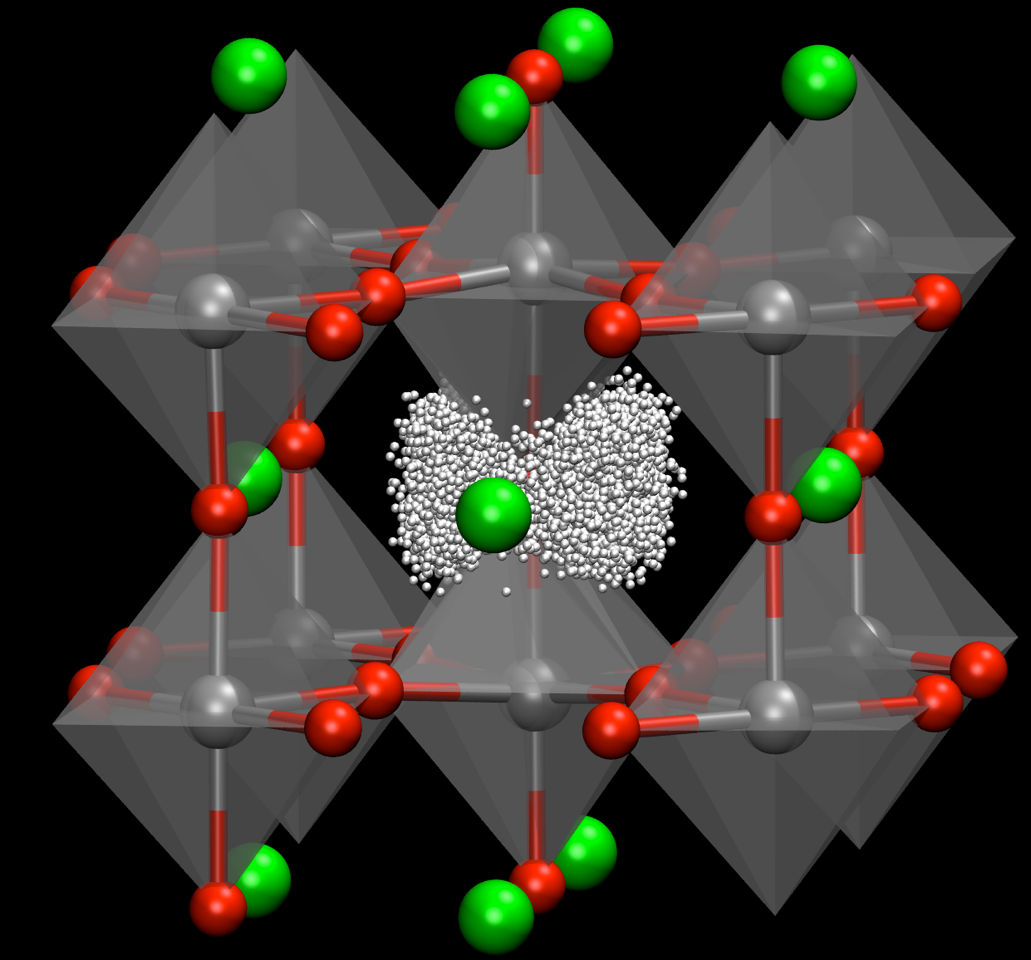} 
   \caption{(Colour online) Positions of a non-reactive proton (small white spheres) sampled during a short simulation at 1500 K. It is evident how the proton can freely rotate around the oxygen to which it is bonded. Here Zr (grey) and O (red) atoms are in the centre and on the vertexes of the octahedra and Ba (green) atoms are in between them.}
   \label{fig:equi}
\end{figure}

\begin{figure}[htbp]
   \centering
   \includegraphics[width=10cm]{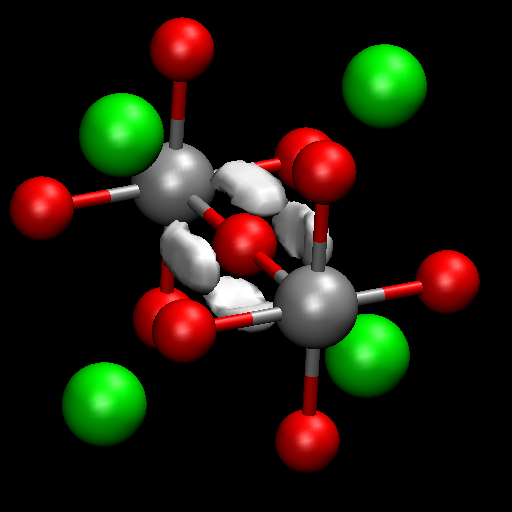} 
   \caption{(Colour online) Probability isosurface for a proton (light grey) during an EVB-MD simulation at 500 K. The isosurface shows a four-fold symmetry around a Zr-O-Zr bond. Ba (green) atoms are also shown as larger light grey spheres without bonds.}
   \label{fig:lobes}
\end{figure}

Having determined a force field representation of the underlying quantum mechanical energy surface, 
it is now possible to perform molecular dynamics simulations of the proton within \bazro. 
Before introducing reactivity into the system, the dynamics of a proton in undoped \bazro ~have been 
examined to check the behaviour of the present model. As shown in Figure \ref{fig:equi}, at 1500 K 
the proton is able to rapidly explore each of the 8 isoenergetic minima accessible to it.
In Figure \ref{fig:lobes} the hydrogen location probability isosurface is also shown for a low temperature run at 500 K. 
Here the eight lobes associated with the hydrogen being bonded to the central 
oxygen are broadened into four elongated surfaces. This indicates that even at lower temperatures
the hopping between different tilt angles of the hydroxyl group occurs rapidly, while the rotation 
about the Zr-O-Zr axis is hindered. 

Activating the Empirical Valence Bond model on top of the underlying potential energy surface allows
us to now explore the diffusivity of hydrogen in \bazro. Here we consider the proton migration in both
pure \bazro ~and in the presence of an yttrium dopant. For the present calculations we have used a 
6x6x6 supercell contain 216 formula units. Note that the force field model has no imaginary modes
anywhere in the Brillouin zone and therefore the use of an even-sized supercell will have no 
repercussions. In the present work we consider just a single Y atom within 
this supercell leading to a concentration of 0.46\%. While experimental studies typically consider much
higher concentrations, as do other simulations, we defer the consideration of multiple dopants and
microstructural complexity to future work. For \bazro, the approximation of examining a single dopant
atom can be justified based on the experimental observation that the activation for diffusion is 
relatively insensitive to the yttrium concentration up 15\% \cite{kreuer03}. 

\begin{figure}[htbp]
   \centering
   \includegraphics[width=10cm]{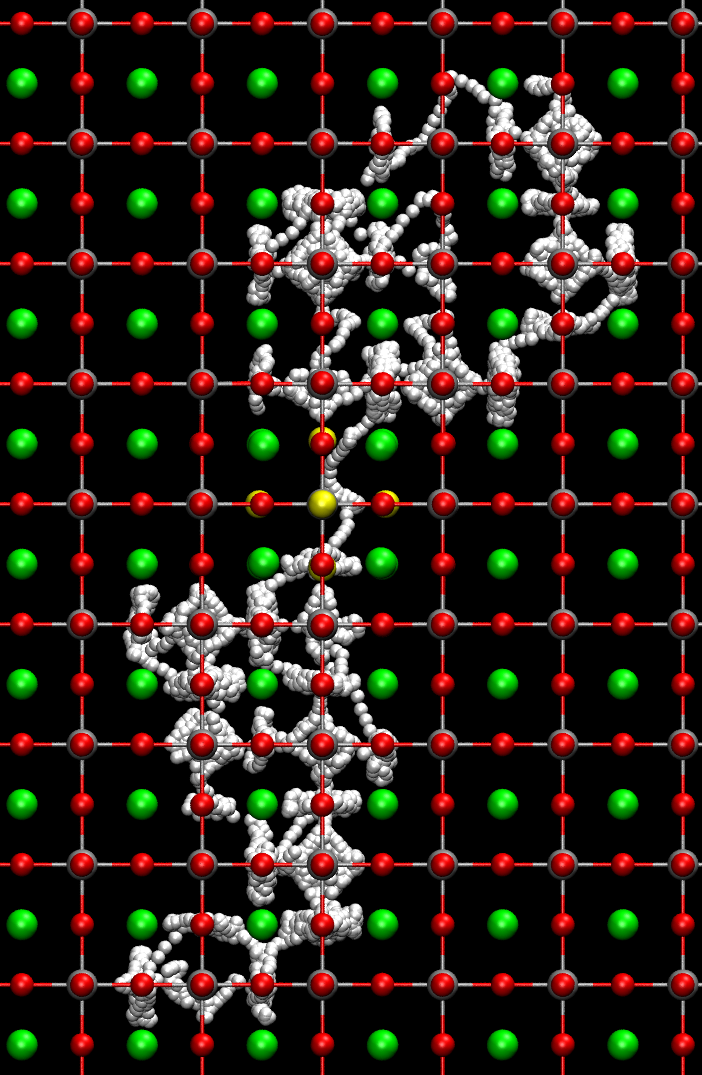} 
   \caption{(Colour online) Diffusion trajectory of a proton (small white spheres) at 1500 K in the BaZrO$_3$ lattice. For clarity the proton trajectory has been smoothed and every sphere represents the proton position averaged over five successive frames taken at 1 ps intervals. The O (red) atoms form square lattice and the Ba (green) atoms are in the centre of the squares. From this view point the Zr atoms are screened by the oxygen atoms at the squares vertexes. In the centre of the image a light grey atom on the square lattice indicates the oxygen that is nearest neighbours of a Y atom underneath.}
   \label{fig:diff2}
\end{figure}

In Figure \ref{fig:diff2}, a sample trajectory for proton diffusion at 1500 K is illustrated. Although it is difficult to 
capture the full dynamics in a static projection, the tendency of the hydrogen to diffuse along the 
edges of octahedra can be seen, as can the high degree of mobility at this temperature. In this particular
trajectory the proton was initially placed at a site remote from the yttrium dopant, but the hydrogen diffuses
past this octahedron during the timescale of the simulation. The low density of time-averaged hydrogen 
positions close to yttrium illustrates that the proton is not trapped in this region, but is able to diffuse by
readily. This is an important feature of the present model. In an earlier version of the model, the use of 
formal charges led to an overestimation of the proton-dopant binding and this resulted in irreversible
trapping on the timescale of the simulations. Through the introduction of a self-energy term and careful
fitting of the model to the quantum mechanical energies for a range of proton binding sites this overbinding
is avoided.

Although the proton is found to be able to diffuse close to yttrium without becoming bound, in some
trajectories trapping is observed. At 1500 K, where the largest number of events are seen and therefore
the statistical reliability is greatest, a proton spends on average 5 ps coordinated to a given oxygen
between two zirconium ions before migrating to an adjacent site. In contrast, when the proton does
become bound to an oxygen in a Y-O-Zr linkage, the proton resides there for between 50 and 90 ps in all
cases. Therefore, the residence time is an order of magnitude greater adjacent to the dopant at this
elevated temperature. At lower temperatures the residence times are longer, as would be expected, but
the longer lifetime of the proton bound adjacent to yttrium persists. 

\begin{figure}[htbp]
   \centering
   \includegraphics[width=10cm]{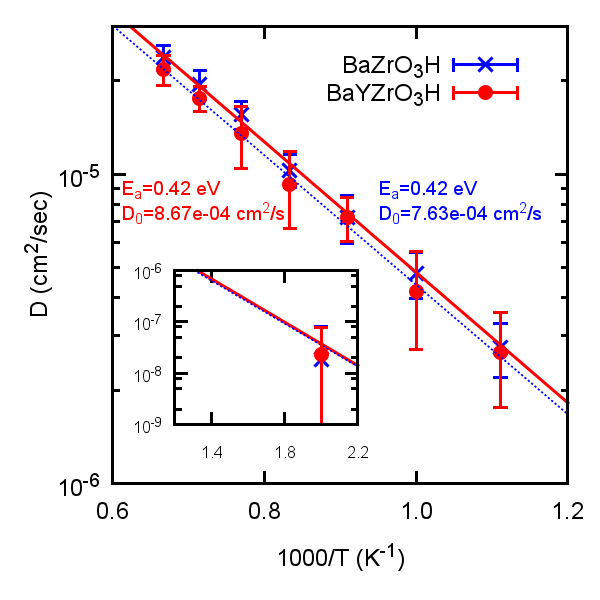} 
   \caption{Proton diffusion coefficient at different temperatures for the pure and Y doped \bazro ~material.
   For clarity the low temperature point (500K) has been placed in the inset.}
   \label{fig:dcoeff}
\end{figure}

In order to quantify the rate of proton migration, we have used an expression for the diffusion coefficient
based on the jump frequency between sites;

\begin{equation}
D=\frac{1}{6} a^2\nu
\end{equation}
where $a$ is the hopping distance and $\nu$ is the average hopping time measured during the simulations. Given that the hydrogen is free to rotate around the octahedra vertices, we assumed the hopping length to be the average oxygen distance. Here we have performed 12 independent simulations for each temperature, where each run samples
1 ns of real time. The values from the ensemble of simulations are used to 
determine the average diffusion coefficient, as shown in Figure \ref{fig:dcoeff}, and the error bar reflects the 
maximum variation
between different individual runs. As the frequency of site jumps increases with temperature, so the 
variability between runs decreases as the temperature rises, leading to a reduced uncertainty. 
The data for 500 K are not included in the fits to determine the activation energies, since only 25\%
of all simulations exhibited any diffusion events. The result for this temperature is shown as an inset
in Figure \ref{fig:dcoeff} in order to demonstrate that the linear fit to the diffusion coefficients does pass within
the range of the average value for this temperature and the upper bound to the observed migration rate.

In Figure \ref{fig:dcoeff}, the results for both doped and undoped \bazro ~are given. As can be seen, the two 
curves are parallel to each other, leading to the same activation energy, and are essentially equivalent 
within the uncertainty. This indicates that the presence of yttrium at low concentration plays little role
in influencing proton diffusivity. This arises primarily because the proton is infrequently trapped adjacent
to  yttrium, and even when this occurs diffusion is only halted for several tens of picoseconds, out of a
nanosecond of simulation, before it escapes. The reduced hydrogen mobility at higher yttrium concentrations
can therefore be postulated to arise from the loss of a pathway through the system that avoids
frequent encounters with yttrium. 

\begin{figure}[htbp]
   \centering
   \includegraphics[width=10cm]{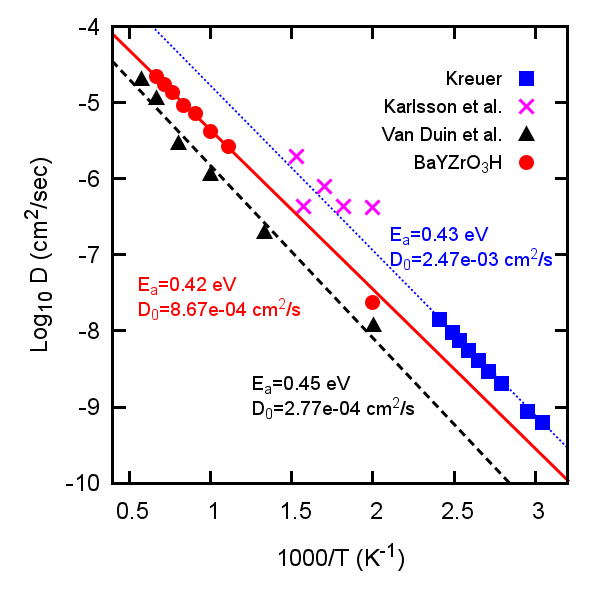} 
   \caption{Comparison between the proton diffusion coefficient calculated in this work using EVB-MD (red dots)
   with the experimental values reported for BaZr$_{0.9}$Y$_{0.1}$O$_{3-\delta}$ by Kreuer \cite{kreuer03}
   (blue squares), Karlsson et al~\cite{Karlsson2010} (purple crosses) and with previous simulation of BaZr$_{0.875}$Y$_{0.125}$H$_{0.125}$O$_3$
   with the ReaxFF force field \cite{vanduin08} (black triangles). }
   \label{fig:dcoeff-exp}
\end{figure}

Finally, we can compare the proton diffusion coefficients obtained in the present work with those of
other simulations with the ReaxFF force field and also against experiment, as shown in Figure \ref{fig:dcoeff-exp}.
Again, we note in comparing results the caveat regarding the different yttrium concentration in the present work. 
This aside, it is found that the activation energy for proton migration is very similar in all cases, with the EVB
method being marginally closer to experiment. Both the ReaxFF and EVB approaches were designed to 
be parametric representations of a predetermined potential energy surface computed using density functional
theory, though with very different underlying functional forms. Hence, it is satisfying that both methods yield
comparable results as it suggests that the quantum mechanical energy surface is being captured properly
overall. 

Where there is a discrepancy between theory and experiment is in the prefactor,
with both molecular dynamics simulations underestimating the experimental value. There are several 
possible reasons for this, including the neglect of quantum effects for the hydrogen nucleus and/or failure
to properly describe the vibrations in key regions of the energy surface. Interestingly, the combined experimental
and theoretical study of Karlsson \textit{et al.} \cite{Karlsson2010} includes first principles estimated 
diffusion coefficients that 
have a prefactor that is too large relative to the measured values. Given the sensitivity of the activation energies
to lattice parameter, it may be that thermal expansion and the consequences for the underlying energy 
surface is the key to the prefactor. In the case of the first principles results the lattice parameter is the 
static one, whereas in the molecular dynamics simulations the unit cell will expand considerably, and 
in all probability exceed
the experimental value due to the lack of quantisation of the vibrations. Further investigation of the influence
of the cell fluctuations, thermal expansion and their correlation with the diffusion coefficient is required to 
determine whether this is a significant factor. 

\section{Conclusions}

A new reactive force field model for the diffusion of hydrogen in the solid oxide fuel cell material, barium
zirconate, has been created based on a potential energy surface generated using density functional 
theory data with the AM05 exchange-correlation functional. By combining a conventional rigid ion 
interatomic potential set with the empirical valence bond approach, the process of deriving the 
force field to describe equilibrium configurations and transition states for hydroxyl group rotation
can be separated from the parameterisation of the proton hop between oxygens along the edge of
a B-site octahedron. 

By using AM05 as the functional, the present model is better able to reproduce the equilibrium structure
of \bazro ~than previous quantum mechanically derived force fields, and the ground state is fitted to be
cubic in accord with experiment \cite{akbar} within the force field. 
Although the computed activation energies for hydroxyl rotation and
proton hopping between two oxygens are quite different to the values from some previous quantum 
mechanical studies, the magnitude of the highest barrier is in line with some of the literature values.
The origin of this overall similarity can, at least partly, be traced back to the variation of the activation energy with
cell parameter and the fact that the barriers for the two key processes, rotation and hopping, change in opposite
directions. For the present method, changing the cell length from 4.0 to 4.4 \AA ~causes the barrier to rotation 
to decrease from 0.51 to 0.17 eV, while the hopping barrier increases from close to zero to 0.31 eV. Hence,
as the cell varies in size there will always be a barrier greater than 0.2 eV, but the character of the rate limiting
process will vary. By fitting a model that accurately captures the volume and using molecular dynamics to 
sample the rate of diffusion, while allowing the cell to fluctuate in an $NPT$ ensemble, this may achieve the
correct balance between different proton motions. This is supported by the finding that the overall activation 
energy, as computed from the temperature dependence of the diffusion coefficient, is in good agreement with experiment. 

Having validated a new reactive model for proton diffusion, it is now possible to consider the broader 
application. Although only a low concentration of yttrium has been examined in simulations so far, the 
extension to higher dopant levels should offer no complications given that the model includes information
on the fully doped limit. Based on the comparison of proton migration in pure and doped
\bazro, albeit at a low concentration so far, the effect of the presence of yttrium is minimal on the 
overall diffusion rate as the increased residence time of the hydrogen near the dopant is limited by
the low probability to become trapped at this site. 

Aside from the consideration of dopant configurations as a function of concentration, the use of a reactive
force field makes it feasible to examine more complex environments, including extended defects, such as
surfaces, dislocations and grain boundaries, though the present model will require more complex 
extension unless the proton is the only reactive species. 
Here systematic enumeration of transition states through the
use of direct quantum mechanical calculations can rapidly become prohibitive. Examination of the contribution
of these features to the macroscopic rate of proton diffusion is now possible in the future. 

\begin{ack}
We thank the Australian Research Council for funding through the Discovery grant DP0986999, as well as iVEC and NCI for providing computing resources.

\end{ack}

\section*{References}

\bibliographystyle{ieeetr}

\begin{thebibliography}{10}

\bibitem{Fabbri10}
E.~Fabbri, D.~Pergolesi, and E.~Traversa {\em Chem. Soc. Rev.}, vol.~39,
  pp.~4355--4369, 2010.

\bibitem{Malavasi10}
L.~Malavasi, C.~A.~J. Fisher, and M.~S. Islam {\em Chem. Soc. Rev.}, vol.~39,
  pp.~4370--4387, 2010.

\bibitem{kreuer03}
K.~D. Kreuer {\em Annu Rev Mater Res}, vol.~33, pp.~333--359, 2003.

\bibitem{munch97}
W.~M\"{u}nch, G.~Seifert, K.~D. Kreuer, and J.~Maier {\em Solid State Ionics},
  vol.~97, pp.~39--44, 1997.

\bibitem{Davies99}
R.~A. Davies, M.~S. Islam, and J.~D. Gale {\em Sol. State Ionics}, vol.~126,
  pp.~323--335, 1999.

\bibitem{BilicGale07}
A.~Bilic and J.~D. Gale {\em Chem. Mater.}, vol.~19, pp.~2842--2851, 2007.

\bibitem{Zhong95}
W.~Zhong and D.~Vanderbilt {\em Phys. Rev. Lett.}, vol.~74, p.~2587, 1995.

\bibitem{Bennett06}
J.~W. Bennett, I.~Grinberg, and A.~M. Rappe {\em Phys. Rev. B}, vol.~73,
  p.~180102, 2006.

\bibitem{BilicGale09}
A.~Bilic and J.~D. Gale {\em Phys. Rev. B}, vol.~79, p.~174107, 2009.

\bibitem{gomez05}
M.~A. Gomez, M.~A. Griffin, S.~Jindal, K.~D. Rule, and V.~R. Cooper {\em J.
  Chem. Phys.}, vol.~123, p.~094703, 2005.

\bibitem{BilicGale08}
A.~Bilic and J.~D. Gale {\em Solid State Ionics}, vol.~179, pp.~871--874, 2008.

\bibitem{gomez10}
M.~A. Gomez, M.~Chunduru, L.~Chigweshe, L.~Foster, S.~J. Fensin, K.~M.
  Fletcher, and L.~E. Fernandez {\em J. Chem. Phys.}, vol.~132, p.~214709,
  2010.

\bibitem{SaulCatlow}
P.~Saul, C.~R.~A. Catlow, and J.~Kendrick {\em Phil. Mag. B}, vol.~51,
  pp.~107--117, 1985.

\bibitem{ZeoPots}
K.-P. Schr\"{o}der and J.~Sauer {\em J. Phys. Chem.}, vol.~100,
  pp.~11043--11049, 1996.

\bibitem{vanduin08}
A.~C.~T. van Duin, B.~V. Merinov, S.~S. Han, C.~O. Dorso, and W.~A. Goddard~III
  {\em J Phys Chem A}, vol.~112, no.~45, pp.~11414--11422, 2008.

\bibitem{ReaxFF}
A.~C.~T. van Duin, S.~Dasgupta, F.~Lorant, and W.~A. Goddard~III {\em J Phys
  Chem A}, vol.~105, pp.~9396--9409, 2001.

\bibitem{GaleWrightRIMG}
J.~D. Gale and K.~Wright {\em Rev. Miner. Geochem.}, vol.~71, pp.~391--411,
  2010.

\bibitem{WarshelWeiss}
A.~Warshel and R.~M. Weiss {\em J. Am. Chem. Soc.}, vol.~102, pp.~6218--6226,
  1980.

\bibitem{MSEVB3}
Y.~Wu, H.~Chen, F.~Wang, F.~Paesani, and G.~A. Voth {\em J. Phys. Chem. B},
  vol.~112, pp.~467--482, 2008.

\bibitem{SIESTA2}
J.~M. Soler, E.~Artacho, J.~D. Gale, A.~Garcia, J.~Junquera, P.~Ordejon, and
  D.~Sanchez-Portal {\em J. Phys.: Condens. Matter}, vol.~14, p.~2745, 2002.

\bibitem{TMpseudo}
N.~Trouiller and J.~L. Martins {\em Phys. Rev. B}, vol.~43, pp.~1993--2006,
  1991.

\bibitem{SIESTA1}
J.~Junquera, O.~Paz, D.~Sanchez-Portal, and E.~Artacho {\em Phys. Rev. B},
  vol.~64, p.~235111, 2001.

\bibitem{MorenoSoler}
J.~Moreno and J.~M. Soler {\em Phys. Rev. B}, vol.~45, pp.~13891--13898, 1992.

\bibitem{AM05}
R.~Armiento and A.~E. Mattsson {\em Phys. Rev. B}, vol.~72, p.~085108, 2005.

\bibitem{Spagnoli10}
D.~Spagnoli, K.~Refson, K.~Wright, and J.~D. Gale {\em Phys. Rev. B}, vol.~81,
  p.~094106, 2010.

\bibitem{AM05water}
A.~E. Mattsson and T.~R. Mattsson {\em J. Chem. Theory Comput.}, vol.~5,
  pp.~887--894, 2009.

\bibitem{GULP97}
J.~D. Gale {\em J. Chem. Soc., Faraday Trans.}, vol.~1, pp.~629--637, 1997.

\bibitem{GULP3}
J.~D. Gale and A.~L. Rohl {\em Mol. Simul.}, vol.~29, pp.~291--341, 2003.

\bibitem{GULPFit}
J.~D. Gale {\em Phil. Mag. B}, vol.~73, pp.~3--19, 1996.

\bibitem{MSEVBborgis}
R.~Vuilleumier and D.~Borgis {\em J. Phys. Chem. B}, vol.~102, pp.~4261--4264,
  1998.

\bibitem{MSEVBmod}
R.~Kumar, R.~Christie, and K.~Jordan {\em J. Phys. Chem. B}, vol.~113,
  pp.~4111--4118, 2009.

\bibitem{MSEVBoh}
C.~D. Wick and L.~X. Dang {\em J. Phys. Chem. A}, vol.~113, pp.~6356--6364,
  2009.

\bibitem{MSEVBoh2}
I.~S. Ulimtsev, A.~G. Kalinichev, T.~J. Martinez, and R.~J. Kirkpatrick {\em
  Phys. Chem. Chem. Phys.}, vol.~11, pp.~9420--9430, 2009.

\bibitem{bussi09jcp}
G.~Bussi, T.~Zykova-Timan, and M.~Parrinello {\em J Chem Phys}, vol.~130,
  no.~7, p.~074101, 2009.

\bibitem{bussi07jcp}
G.~Bussi, D.~Donadio, and M.~Parrinello {\em J Chem Phys}, vol.~126, no.~1,
  p.~014101, 2007.

\bibitem{bussi08cpc}
G.~Bussi and M.~Parrinello {\em Computer Physics Communications}, vol.~179,
  pp.~26--29, Jul 2008.

\bibitem{parrinello80prl}
M.~Parrinello and A.~Rahman {\em Phys. Rev. Lett.}, vol.~45, no.~14,
  pp.~1196--1199, 1980.

\bibitem{parrinello81jap}
M.~Parrinello and A.~Rahman {\em Journal of Applied Physics}, vol.~52, p.~7182,
  1981.

\bibitem{martyna94jcp}
G.~Martyna, D.~Tobias, and M.~L. Klein {\em J Chem Phys}, vol.~101, p.~4177,
  1994.

\bibitem{trotter1959}
H.~F. Trotter {\em Proceedings of the American Mathematical Society}, vol.~10,
  pp.~545--551, 1959.

\bibitem{tuckerman92jcp}
M.~E. Tuckerman, B.~Berne, and G.~J. Martyna {\em J. Chem. Phys.}, vol.~97,
  no.~3, pp.~1990--2001, 1992.

\bibitem{crank47}
J.~Crank and P.~Nicolson {\em Proc. Camb. Phil. Soc.}, vol.~43, p.~50Ð67, 1947.

\bibitem{Bussi2007pre}
G.~Bussi and M.~Parrinello {\em Phys. Rev. E}, vol.~75, no.~5, p.~056707, 2007.

\bibitem{bjoerketun07}
M.~E. Bj\"{o}rketun, P.~G. Sundell, and G.~Wahnstr\"{o}m {\em Phys. Rev. B},
  vol.~76, p.~054307, 2007.

\bibitem{pw91}
J.~P. Perdew, J.~A. Chevary, S.~H. Vosko, K.~A. Jackson, M.~R. Pederson, D.~J.
  Singh, and C.~Fiolhais {\em Phys. Rev. B}, vol.~46, pp.~6671--6687, 1992.

\bibitem{Merinov09}
B.~Merinov and W.~A. Goddard~III {\em J. Chem. Phys.}, vol.~130, p.~194707,
  2009.

\bibitem{pbe}
J.~P. Perdew, K.~Burke, and M.~Ernzerhof {\em Phys. Rev. Lett.}, vol.~77,
  p.~3865, 1996.

\bibitem{pbepw91differ}
A.~E. Mattsson, R.~Armiento, P.~A. Schultz, and T.~R. Mattsson {\em Phys. Rev.
  B}, vol.~73, p.~195123, 2006.

\bibitem{stokes10}
S.~J. Stokes and M.~S. Islam {\em J. Mater. Chem.}, vol.~20, pp.~6258--6264,
  2010.

\bibitem{Karlsson2010}
M.~Karlsson, D.~Engberg, M.~E. Bj\"{o}rketun, A.~Matic, G.~Wahnstr\"{o}m, P.~G.
  Sundell, P.~Berastegui, I.~Ahmed, P.~Falus, B.~Farago, L.~B\"{o}rjesson, and
  S.~Eriksson {\em Chem. Mater.}, vol.~22, pp.~740--742, 2010.

\bibitem{ParrinelloElastic}
M.~Parrinello and A.~Rahman {\em J. Chem. Phys.}, vol.~76, pp.~2662--2666,
  1982.

\bibitem{akbar}
A.~R. Akbarzadeh, I.~Kornev, C.~Malibert, L.~Bellaiche, and J.~M. Kiat {\em
  Phys. Rev. B}, vol.~72, p.~205104, 2005.

\bibitem{candg}
Z.~P. Chang and E.~K. Graham {\em J. Phys. Chem. Solids}, vol.~38,
  pp.~1355--1362, 1977.

\bibitem{yamanaka}
S.~Yamanaka, M.~Fujikane, T.~Hamaguchi, H.~Muta, T.~Oyama, T.~Matsuda,
  S.~Kobayashi, and K.~Kurosaki {\em J. Alloys Comp.}, vol.~359, pp.~109--113,
  2003.

\bibitem{goretta}
K.~C. Goretta, E.~T. Park, R.~E. Koritala, M.~M. Cuber, E.~A. Pascual, N.~Chen,
  A.~R. de~Arellano-Lopez, and J.~L. Routbort {\em Physica C}, vol.~309,
  pp.~245--250, 1998.

\end{thebibliography}

\clearpage
\begin{table}[htbp]
   \centering
   \begin{tabular}{|c|c|c|ccc|}
   \hline
   Material & Property & Experiment & AM05 & ReaxFF & EVB \\
   \hline
   BaO & $a$ (\AA)       & 5.5391 & 5.5207 & 5.5987 & 5.5207 \\
     & $C_{11}$ (GPa)& 124.4 \cite{candg} &  & 87 & 121 \\
     & $C_{44}$ (GPa)&  48.7 \cite{candg}   &  & 32 &  58 \\	
     & $C_{12}$ (GPa)&  33.7 \cite{candg}  &  & 40 &  58 \\
   \hline
   \bazro & $a$ (\AA) & 4.191/4.192 \cite{akbar,yamanaka} & 4.1898 & 4.2658 & 4.1898 \\
     & Bulk modulus (GPa)  & 127 \cite{yamanaka} & & 90 & 214 \\
     & Shear modulus (GPa) & 97/103 \cite{goretta,yamanaka} & & 59 & 128 \\
     & Young's modulus (GPa) & 241/243 \cite{goretta,yamanaka} & & 149 & 299 \\
     & Poisson's ratio (GPa) & 0.237 \cite{goretta} & & 0.225 & 0.267 \\
   \hline
   \end{tabular}
   \caption{Comparison of structure and properties obtained for BaO and \bazro ~as measured experimentally
   and calculated with two force fields,
   namely the ReaxFF model and that of the present work fitted to the quantum mechanical data. In
   the case of the ReaxFF force field values are given for the unstable cubic \bazro ~structure. The
   optimised cell parameters computed using density functional theory with the AM05 functional are
   also given. }
   \label{bulk_results}
\end{table}

\clearpage
\begin{table}[htbp]
   \centering
   \begin{tabular}{|c|c|c|c|}
   \hline
   Configuration & r(Y-H) (\AA) & $\Delta U_{AM05}$ (eV) & $\Delta U_{EVB}$ (eV) \\
   \hline
   1 & 2.179  &  0.000 &  0.000  \\
   2 & 3.110  &  0.061 &  0.116  \\
   3 & 3.643  &  0.006 &  0.047  \\
   4 & 5.584  &  0.129 &  0.124  \\
   5 & 8.979  &  0.237 &  0.238  \\
   \hline
   \end{tabular}
   \caption{Energies of the different configurations for proton in Y-doped \bazro,
   taken relative to the most stable state. Both the values given by the quantum mechanical
   calculations with the AM05 functional and by the resulting EVB force field fit are reported.
   The minimum yttrium-hydrogen distance is also given for the optimised quantum mechanical
   geometry in a 3x3x3 supercell as an indication of the proton location.}
   \label{yoh_energies}
\end{table}

\clearpage
\centering
\begin{table}
\caption{Intermolecular force field parameters for the rigid ion model derived in the present work. Here the atom types O1, O2 and O3 distinguish between the oxygens of a hydroxyl group, an oxide ion bonded to Zr only, and an oxide ion bond to Y and Zr, respectively. Where just O is given this implies that the potential acts between all types of oxygen. Formal charges are used for all species, except H and O1, which have charges of 0.308698 and -1.308698 a.u., respectively. All intermolecular interactions are of the form of the Buckingham potential, except for the intermolecular H-O1 interaction, which is given by a Lennard-Jones 12-6 potential. The intramolecular H-O1 interaction is represented via a Coulomb-subtracted harmonic potential with a force constant and equilibrium distance of 46.016 eV \AA$^{-2}$ and 0.985357 \AA, respectively. Interactions are smoothly tapered to zero over a range of 2 \AA, starting at 8 \AA.
  }
\centering
   \label{forcefield}
  \begin{tabular}{cccccc}
\\
\hline
\multicolumn{2}{c}{Buckingham}
               & A[eV]        & $\rho$[\AA]\\
\hline
Ba  &    O1     & 688.2166    &  0.392014  \\
Ba  &    O2     & 972.1995   &  0.392014  \\
Ba  &    O3     & 972.1995   &  0.392014  \\
Zr  &    O1     & 2195.8712    &  0.298760 \\
Zr  &    O2     & 3250.8545   &  0.298760  \\
Zr  &    O3     & 3250.8545   &  0.298760  \\
Y   &    O1     & 2972.3808    &  0.267317  \\
Y   &    O2     & 4121.8373   &  0.280386  \\
Y   &    O3     & 4121.8373   &  0.280386  \\
O    &    O     & 22764.0    &  0.149  \\
\hline 
\multicolumn{2}{c}{Lennard--Jones}
               &  A [eV\AA$^{12}$]   \\
\hline
H1    &   O    & 6.391    \\
H1    &   O3   & 7.800   \\
\hline
\end{tabular}
\end{table}

\clearpage
\begin{table}[htbp]
   \centering
   \begin{tabular}{|c|l|l|c|c|}
   \hline
   & Stochastic & Langevin & Static (0K) & Static (300K) \\
   \hline
   $C_{11}$ (GPa)& 361 $\pm$ 15 & 356 $\pm$ 7 & 371.1 & 365.7 \\
   $C_{44}$ (GPa)& 137 $\pm$ 7  & 134 $\pm$ 3 & 135.3 & 134.9 \\	
   $C_{12}$ (GPa)& 131 $\pm$ 15 & 139 $\pm$ 6 & 135.3 & 134.0 \\
   \hline
   \end{tabular}
   \caption{Average elastic calculated from the cell fluctuations during an $NST$ run at 300K
   with $\tau_b$=0.1 ps
   using different $\tau_t$ and $\zeta$ for the Stochastic and Langevin thermostats.
   The values obtained using lattice dynamics (static) with GULP are also shown for comparison.
   Static values are calculated using the optimised lattice parameters (0 K) and the average cell
   dimensions from the molecular dynamics (300 K).}
   \label{tab:booktabs}
\end{table}

\clearpage

\clearpage

\clearpage

\clearpage

\clearpage

\clearpage

\end{document}